\documentclass[aps,pra,twocolumn,superscriptaddress]{revtex4}
\usepackage{natbib}
\usepackage{amsmath,amssymb}

\usepackage{graphicx}
\usepackage[utf8]{inputenc}

\newcommand{\fisicarm}{Dipartimento di Fisica, Sapienza Universit\`{a} di Roma, Piazzale Aldo Moro, 5, I-00185 Roma, Italy}
\newcommand{\fisicami}{Dipartimento di Fisica, Politecnico di Milano, Piazza Leonardo da Vinci, 32, I-20133 Milano, Italy}
\newcommand{\pisasns}{NEST, Scuola Normale Superiore and Istituto di Nanoscienze - CNR, I-56126 Pisa, Italy}
\newcommand{\ino}{Istituto Nazionale di Ottica, Consiglio Nazionale delle Ricerche (INO-CNR), Largo Enrico Fermi, 6, I-50125 Firenze, Italy}
\newcommand{\ifn}{Istituto di Fotonica e Nanotecnologie, Consiglio Nazionale delle Ricerche (IFN-CNR), Piazza Leonardo da Vinci, 32, I-20133 Milano, Italy}
\newcommand{\singapore}{Center for Quantum Technologies, National University of Singapore, 117542 Singapore, Singapore}

\begin{document}
\title{Quantum simulation of bosonic-fermionic non-interacting particles in disordered systems via quantum walk}

\author{Francesco De Nicola}
\affiliation{\fisicarm}

\author{Linda Sansoni}
\email{linda.sansoni@gmail.com}
\affiliation{\fisicarm}

\author{Andrea Crespi}
\affiliation{\ifn}
\affiliation{\fisicami}

\author{Roberta Ramponi}
\affiliation{\ifn}
\affiliation{\fisicami}

\author{Roberto Osellame}
\affiliation{\ifn}
\affiliation{\fisicami}

\author{Vittorio Giovannetti}
\affiliation{\pisasns}

\author{Rosario Fazio}
\affiliation{\pisasns}
\affiliation{\singapore}

\author{Paolo Mataloni}
\affiliation{\fisicarm}
\affiliation{\ino}

\author{Fabio Sciarrino}
\affiliation{\fisicarm}
\affiliation{\ino}

\begin{abstract}

We report on the theoretical analysis of  bosonic and fermionic non-interacting systems in a discrete two-particle quantum walk affected by different kinds of disorder. We considered up to 100-step QWs with a spatial, temporal and space-temporal disorder observing how the randomness and the wavefunction symmetry non-trivially affect the final spatial probability distribution, the transport properties and the Shannon entropy of the walkers.
\end{abstract}
\maketitle

In statistical physics random walks  describe the propagation of a particle (the walker) under the action of a probabilistic process which forces the latter to move along preassigned directions (say one step on the left or one step on the right if the system is one dimensional). Despite their  simplicity, random walks have found applications in many research fields, spanning from economics, computer science, chemistry, and physics. A quantum version of this model was first provided by Y. Aharonov,  L. Davidovich and N. Zagury \citep{ahar93pra} who introduced the notion of {\em quantum walks}.\\
Unlike {their classical counterparts}, in quantum walks, the final state of the walker is highly sensitive to the initial conditions of the system. Indeed, during its evolution, the particle  spatial distribution does not converge to a steady state, but it spreads ballistically. After few steps, the counterintuitive profile of the wavefunction emerges as a result of quantum interference among many possible paths. {The massive parallelism in exploring multiple trajectories, is at the base to simulate} 
biological \citep{Lloyd2008,plen08njp}, chemical \citep{Lloyd2008} and physical \citep{Sansoni2011,OBrien2010,Silberhorn2010} systems and paving the way for universal quantum computation \citep{chil09prl,Childs2012}.
\\
\indent
{Regarding the experimental implementation of} quantum walks, {they} have been observed {in several platforms, such as} nuclear magnetic resonance \citep{Ryan2005}, trapped ions and trapped cold neutral atoms \citep{Schmitz2009,Zahringer2010}, single photons in bulk \citep{P.K.Pathak2007}, fiber optics \citep{Silberhorn2010,Silberhorn2011} and coupled waveguide arrays \citep{Sansoni2011,OBrien2010,Silberberg2009}.
\\
{The nature of propagation of multiple particles in a quantum walk may be strongly affected even in the absence of a direct interaction 
between them. Quantum (non-local) correlations, present in the initial state will influence the overall wave-function evolution  as the
bunching/antibunching observed in interferometry. Entangled walkers have been studied both theoretically~\citep{omar06pra,P.K.Pathak2007,
Silberberg2009,Silberberg2010} and experimentally~\citep{OBrien2010,Sansoni2011,cres13npo} in orderered and disordered systems.
Depending on the symmetry of the input entangled state  it is possible to simulate the particles obeying different (boson/fermion) statistics. 

With reference to the experimental realization of QWs in presence of disorder presented in~\citep{cres13npo}, here we report a detailed 
theoretical and numerical description of propagation of non-interacting  bosonic and fermionic particles in a disordered environment}.
In this photonic approach, walkers are represented by photon pairs sharing polarization entanglement on a QW circuit implemented by an array of cascaded beam splitters, as we will be describe in the following. Generally, photons are limited to the behavior dictated by Bose-Einstein statistics, which defines quantum interference and quantum gates \citep{C.K.Hong1987}. The ability to simulate non-bosonic statistics with photons by means of polarization entanglement, could give access to phenomena otherwise not physically accessible or that would be hidden by decoherence, providing a way to verify quantum simulations performed in other quantum systems \citep{Sansoni2010,Sansoni2011,cres13npo}. Besides that, the controlled engineering of disorder would enable a detailed {understanding of the distinct signatures of statistics  on the system localisation dynamics \citep{cres13npo}.}

{This paper is organised  as follows}. In the first section we briefly review the discrete-time quantum walk and the two-particle probability distributions for an ordered structure. In sections \S\ref{sec:staticdisorderandandersonlocalization}, \S\ref{sec:dynamicdisorderanddecoherence} and \S\ref{sec:fluctuatingdisorderandandersontransition} we analyze different {types of disorder and the effects introduced into the walk  by varying the symmetry of the input states. 
In section \S\ref{sec:Disorderstrength}, we investigate how the disorder strength affects the width of the wave-packet, showing 
how the amount of disorder in the system can be varied in a controlled fashion. Fractality, anomalous diffusion and other transport
 properties of bosonic and fermionic particles are discussed in section \S\ref{sec:Transportproperties}, while Shannon entropy 
 and mutual information of these states are shown in section \S\ref{sec:Shannon entropy}. Section \S\ref{sec:conclusions} is 
 devoted to the conclusions.}   
\section{Discrete-time quantum walk}
\label{sec:DiscreteTimeQuantumWalk}
{{Let us give a brief review of some basic concepts on discrete-time quantum walk. A quantum walker is a quantum particle -such as an electron, atom or photon- characterized by both ``external"  and ``internal" degrees of freedom,
the former describing the propagation of the system in real space, the latter describing the state of the ``coin" that
dictates which path the particle has to follow during the evolution. Accordingly, restricting to  walks along a discrete one-dimensional lattice,  a generic state of the system can be expressed as 
\begin{eqnarray}
	|\Psi \rangle &=& \sum_{x} |\psi(x)\rangle  \otimes\left|x\right\rangle\;, \\
	|\psi(x)\rangle &=&\alpha(x) \left|L\right\rangle+\beta(x) \left|R\right\rangle \;, 
\end{eqnarray}
where $|x\rangle$ defines the  particle position, with  
 $\left\{\left|L\right\rangle,\,\left|R\right\rangle\right\}$ an orthonormal  basis for the coin space (assumed hereafter to be bi-dimensional), and where 
 $\alpha(x),\,\beta(x)$ are complex amplitudes (the probability of finding the particle at position $x$ irrespectively from its internal state being {$P(x) = \langle x |\psi(x)\rangle$}~\citep{Meyer1997}.
In this setting
the dynamics is described as a stroboscopic process which, after $t$ steps, brings a generic 
initial state $|\Psi\rangle$ into the state 
\begin{eqnarray}\label{STROBO}
	|\Psi(t)\rangle&=& \sum_x |\psi(x,t)\rangle \otimes |x\rangle =  \hat{U}^{t} |\Psi\rangle \;,
\end{eqnarray}
 with $\hat{U}=\hat{S}\cdot(\hat{C}\otimes\hat{I})$ being the unitary transformation 
 obtained by first performing 
  an Hadamard transformation  $\hat{C}$ (coin) on the internal degree of freedom followed by a conditional  displacement operator $\hat{S}$ \citep{Kempe2003}, i.e. 
\begin{equation}
\hat{C}=\frac{1}{\sqrt{2}}\left(
\begin{array}{lr}
1 &1 \\
1 &-1
\end{array}
\right)\;,
\end{equation}
\begin{equation}
\hat{S}=\sum_{x}\left|L\right\rangle\left\langle L\right|\otimes \left|x-1\right\rangle\left\langle x\right|
+\left|R\right\rangle\left\langle R\right|
\otimes \left|x+1\right\rangle\left\langle x\right|.
\label{stepOp}
\end{equation}
Assume hence that the quantum walker is initially localized at  position $x=0$ with internal state 
 $|L\rangle$ or  $|R\rangle$.
  As result of quantum interference among multiple paths that originates from $t$ steps
  of the evolution~(\ref{STROBO}),
the counterintuitive profile of the quantum walk probability distribution can be retrieved by
measuring the position of the particle along the 1-D line irrespectively from its internal state, i.e. 
by looking at the quantity  {$P(x;t) = \langle x | \psi(x,t)\rangle$. } 
 
More complex probability distributions arise when two or more particles are injected into the same quantum walk.
 In particular an interesting situation is given by the evolution of identical particles obeying bosonic
 or fermionic statistics. 
In this case, due to  {the symmetrisation postulate of quantum mechanics, (bunching/anti-bunching) is 
expected to influence the dynamics of the quantum walks.}
These effects have been analyzed exploiting all-optical implementations 
of the 
discrete quantum walk dynamics~(\ref{STROBO})~\citep{Sansoni2011,cres13npo}.
In these setups, building up from  the theoretical proposals of \citep{jeon04pra,rohd11njp}
the propagation of a single quantum walker along a 1-D line is simulated with a single photon which
undergoes to multiple scattering from  $50/50$  beam-splitters organized in a bi-dimensional cascade array (see Fig. \ref{fig:BSwalk}).
\begin{figure}[ht]
	\centering
	\includegraphics[width=\columnwidth]{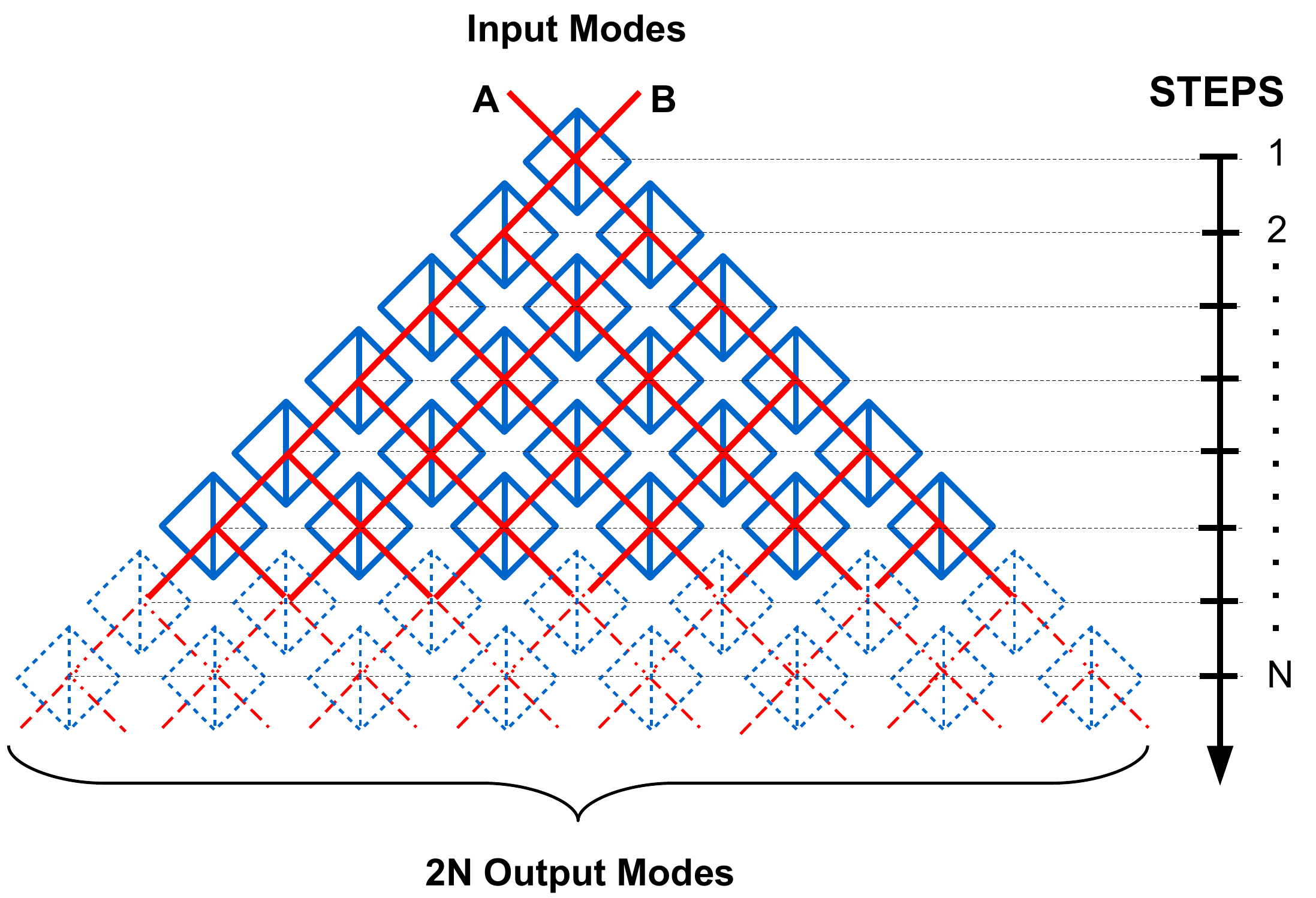}
	\caption{Photonic implementation of 1-D discrete-time quantum walk. Each site is represented by a beam splitter (blue empty box): a photon impinging on a symmetric beam splitter has the same probability to emerge from one of its two outputs. Due to this feature, the beam splitter may be used in a QW as both coin and step operator: by arranging many beam splitters in a cascaded configuration, it is possible to simulate a N-step QW circuit in which each line of beam splitters simulates a step of the QW \citep{Sansoni2011,cres13npo}.}
	\label{fig:BSwalk}
\end{figure}

This is particularly convenient since the conditional 
displacement~(\ref{stepOp})  is automatically implemented (via a dual-rail encoding based on a 
``which-path'' information) in the propagation along the network. Most interestingly in these systems it is possible to emulate  statistical effects associated with the evolution of two indistinguishable particles,  by exploiting the polarization degree of freedom of a couple of photons  to enforce  the proper symmetric/anti-symmetric 
distributions on the effective 1-D line \citep{omar06pra,Sansoni2011,cres13npo}. Specifically, the quantum  walk of a 
pair of bosonic/fermionic particles is obtained by  injecting into the system the symmetric (anti-symmetric) state 
{$|\Psi^{(\pm)}\rangle$}  defined by 
\begin{eqnarray}\label{INPUT1}
|\Psi^{(\pm)} \rangle = \frac{1}{\sqrt{2}} \; \Big( |\psi_A, H\rangle |\psi_B, V\rangle \pm  |\psi_B,H\rangle|\psi_A,V\rangle \Big)\;,
\end{eqnarray} 
where  $|\psi_{A}, H\rangle = |\psi_A\rangle \otimes |H\rangle$ and $|\psi_{B}, H\rangle= |\psi_B\rangle\otimes |H\rangle$ are orthonormal vectors 
 describing  respectively a photon with horizontal polarization which is
 entering the array from two distinct ports of the setup -Fig. \ref{fig:BSwalk}- (similar definitions apply for  $|\psi_{A,B},V\rangle$ associated with a vertically polarized single photon). Assuming that the propagation through the network
 is polarization insensitive and that no interaction is present among the two photons, after $t$ steps  Eq.~(\ref{INPUT1}) will evolve into
 \begin{eqnarray}\label{INPUT2}
|\Psi^{(\pm)} (t) \rangle &=&  \frac{1}{\sqrt{2}} \; \Big( |\psi_A(t), H\rangle |\psi_B(t), V\rangle \nonumber \\
&&\qquad \pm  |\psi_B(t),H\rangle|\psi_A(t),V\rangle \Big)\;,
\end{eqnarray} 
  where for $C=A,B$, $|\psi_C(t)\rangle = U^t | \psi_C\rangle$   is the 
   the  evolved counterpart of the
  input state $ |\psi_{C}\rangle$.
  Fig. \ref{fig:qwordinato} reports the joint probability distribution $P^{(\pm, sym)}(x,y,t)$   
  associated with the 
 detection of  a  photon in position $x$ and the other in position $y$, irrespectively to their
  polarization, computed at $t= 50$ and assuming that $A$ and $B$ are two neighboring input ports of the
  setup (see   Eq.~(\ref{defiSYM}) of the appendix for a formal definition of this quantity). 
 The different effects of symmetric ($P^{(+,sym)}$) and antisymmetric ($P^{(-,sym)}$) distributions, which can be related to bosonic and fermionic statistics respectively, are shown in the corresponding probability distributions and density plots. The anti-bunching feature in Fig. \ref{fig:qwordinato}c,d exhibits zero-probability diagonal elements, meaning that fermions are non-trivially arranged in space \citep{Silberberg2010,Sansoni2011}. Most significantly, this pattern survives the random scattering process even after very long evolution times. While bosons have non-vanishing probability to be at the same site, fermions have zero probability to be at the same site, as expected due to the Pauli exclusion principle \citep{Sansoni2010,C.K.Hong1987}. This profound difference between the two quantum walks is a signature of the probability distribution sensitiveness to the initial state, due to the unitarity of the evolution.}}
\begin{figure}[ht]
	\centering
	\includegraphics[width=\columnwidth]{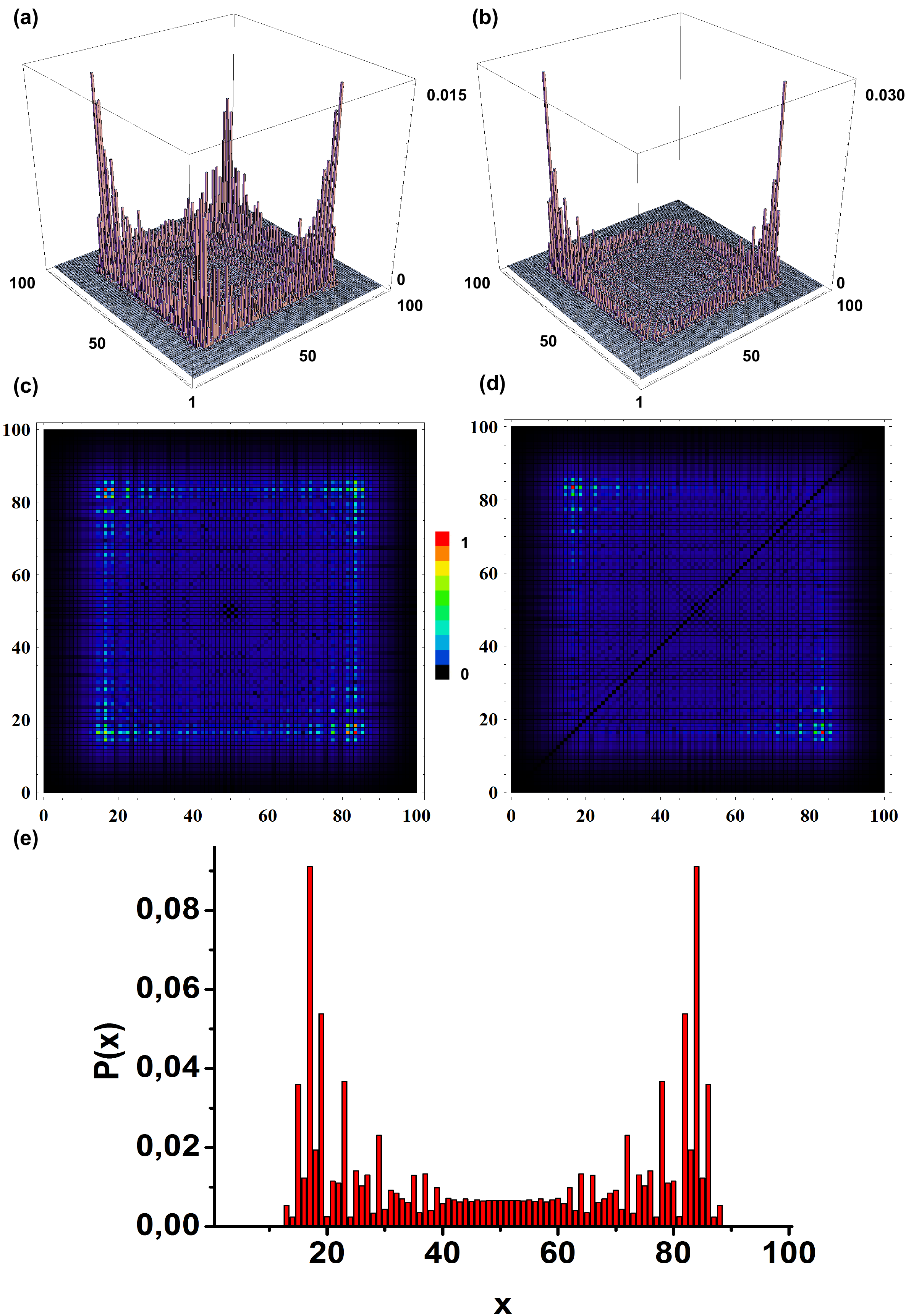}
	\caption{Modes probability distributions $P^{(\pm,sym)}(x,y,t)$ (a,b) and associated density plots (c,d) of bosonic (left) and fermionic (right) two-particle states. At $t=0$, two particles are placed at two neighboring sites in the center of the lattice. The distribution is calculated after $t=50$ steps. The matrices represent the probability to find one particle at the output mode $i$ and one at the output mode $j$. Both bosons and fermions diffuse ballistically giving rise to significant contributions in the corners of the probability distributions. {Similar plots but for $N=30$	steps can be found in Ref. \citep{omar06pra}. (e) Marginal distribution $P^{sym}(x)$ of Eq. (\ref{marginal}) obtained by summing over the columns of plots (c,d). The same distribution arises from both symmetric and anti-symmetric distribution (See Appendix A).}}
	\label{fig:qwordinato}
\end{figure}

\section{Static disorder and Anderson localization}
\label{sec:staticdisorderandandersonlocalization}
\begin{figure}[ht]
	\centering
		\includegraphics[width=\columnwidth]{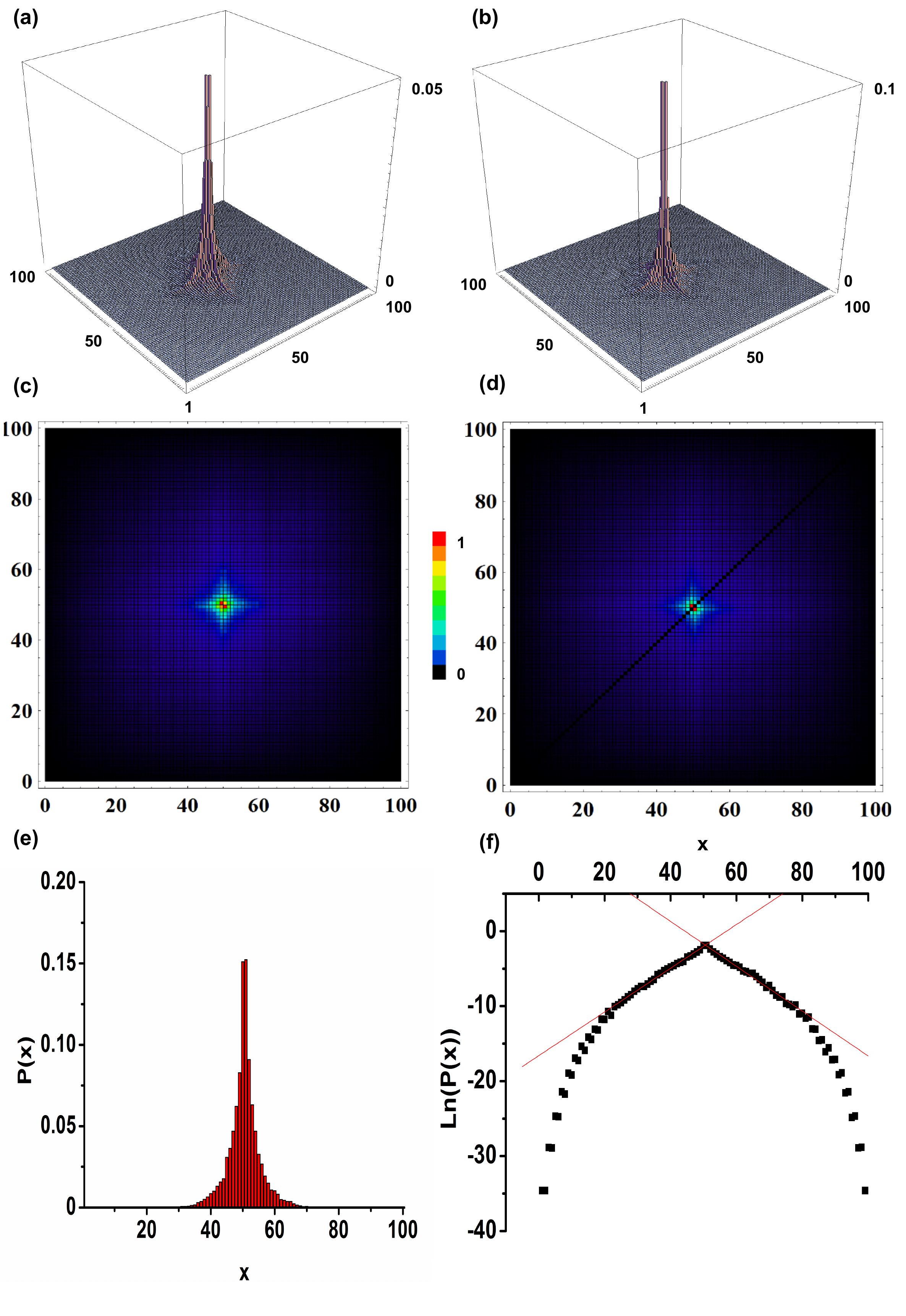}
	\caption{Modes probability distributions (a,b) and associated density plots (b,c) of bosonic (left) and fermionic (right) two-particle states in the case of a quantum walk in presence of static disorder. At $t=0$, two particles are placed at two neighboring sites in the center of the lattice. The distribution is calculated after $50$ steps and averaged over $100$ configurations of static disorder. Both bosons and fermions localize nearby their initial position with a typical distribution profile characterized by an exponential decay. The dependence on the symmetry of the initial state is still observable. {(e) Marginal distribution $P^{sym}(x)$ of Eq. (\ref{marginal}) obtained by summing over the columns of plots (c,d).} Linear fit in semilog scale (f). The localization lenght after $50$ steps of a two-particle quantum walk is estimated to be $\xi\cong3$}	\label{fig:qwstatico}
\end{figure}

{Let us now move to the description of such a scenario in presence of disorder.\\}
Disorder exists at different levels in nature, ranging from a few impurities or interstitial defects in an otherwise perfect crystalline host \citep{Castellani1983,Anderson1958,Mott1967} to the strongly disordered limit of glass structures \citep{Lee1985}.  {In particular, the static 
disorder  is set when a system parameter is a random variable evolving in space but 
not in time~\citep{Lifshits1988}. Such systems exhibit Anderson localization \citep{Anderson1958}.} This phenomenon, has a fundamental relevance since it offers a mechanism to understand, for example, the electronic conductance in imperfect crystals and the metal-insulator transition \citep{Castellani1983,Mott1967,Lee1985}.\\
\indent
As predicted by Anderson, static disorder may lead to the absence of diffusion and the particle wavefunction results localized. In one dimension, it can be rigorously shown that all states are localized because of the repeated backscattering and the localization lenght is proportional to the mean free path of the particle, no matter how strong or weak the disorder is \citep{Mott1967}.\\
\indent
The key factor for the interference effect responsible for the localization is the broken periodicity in the dynamics of the system, induced by the disordered media: random perturbations to the dynamics of the system can break the periodicity, and manifest localization \citep{Silberhorn2011,Chandrashekar2011}. {Localization of the walker} in one dimension may be controlled by introducing drifts with constant momentum between two consecutive steps of the walk, i.e. by applying controlled phase shift operations on the particle wavefunction to randomly stop at each site the evolution of the quantum coin \citep{Chandrashekar2011a}, whose matrix now reads:
\begin{equation}
C(x)=\frac{1}{\sqrt{2}}\left(
	\begin{array}{lr}
	e^{i\phi_{L}(x)} &0 \\
	0 &e^{i\phi_{R}(x)}
	\end{array}
	\right)\left(
\begin{array}{lr}
1 &1 \\
1 &-1
\end{array}
\right).
\end{equation}
At each step, coin operations evolve the initial state by applying the same phases $\left\{\phi_{L}(x),\,\phi_{R}(x)\right\}$. While in absence of disorder quantum walk exhibits a spatial probability distribution diffusing ballistically, in case of static disorder the distribution appears localized with a shape characterized by an exponential decay.\\ 
\indent 
Single particle transport in disordered lattices has been experimentally observed in different frameworks (microwaves in strongly scattering samples \citep{Chabanov2000}, single photons in bulk \citep{Hilke2009}, photonic lattices \citep{Saleh2011}, ultrasounds waves in a three-dimensional elastic system \citep{Faez2009} and Bose-Einstein condensates \citep{Skipetrov2008}); conversely, there are few numerical studies \citep{Silberberg2010, Silberhorn2010} and only one experiment about multi-particle transport \citep{cres13npo}.\\
\indent
Understanding the behavior of a multi-particle system in a disordered medium may be of great interest. Indeed, the bosonic or fermionic nature of the particles strongly affects the transport phenomenon, deriving from both wave interference and quantum correlations. Because of the coin operations unitarity, the static disorder only affects the shape of the distribution and does not alter the symmetry of the initial state. Therefore the quantum walk with static disorder still results strongly dependent on the input state, as the  anti-bunching feature shows with the diagonal-vanishing pattern in Fig. \ref{fig:qwstatico}d.\\
\indent
By tracing out the position of one of the particles (i.e. summing over the columns of the correlation matrix as detailed in Eq. (\ref{marginal})), it is possible to confirm the exponential decay of the Anderson peak by computing the linear fit of the distribution in semilogarithmic scale (Fig. \ref{fig:qwstatico}e,f) and therefore calculating the localization length in inverse proportion to the angular coefficient.

The analysis here reported, dealing with a discrete time quantum walk, could be performed also by exploiting a continuous time QW, indeed the two approaches have shown many similarities \citep{stra06pra}. 
Simulations of space-dependent disorder for two-particle systems have been reported in this case by Lahini et al. \citep{Silberberg2010}, where the two-particle correlation function is calculated after short evolution times, in such a way that each particle has nonzero
probability to be localized or to remain ballistic: it may be associated to an evolution in presence of static disorder whose strength is non-maximal. At variance with this approach, in our scenario we implement a maximal strength static disorder. In this case the probability of the particles to diffuse ballistically becomes negligible on a 100-step QW, allowing us to observe a pure localization effect.

\section{Space correlated dynamic disorder and decoherence}
\label{sec:dynamicdisorderanddecoherence}
\begin{figure}[ht]
	\centering
		\includegraphics[width=\columnwidth]{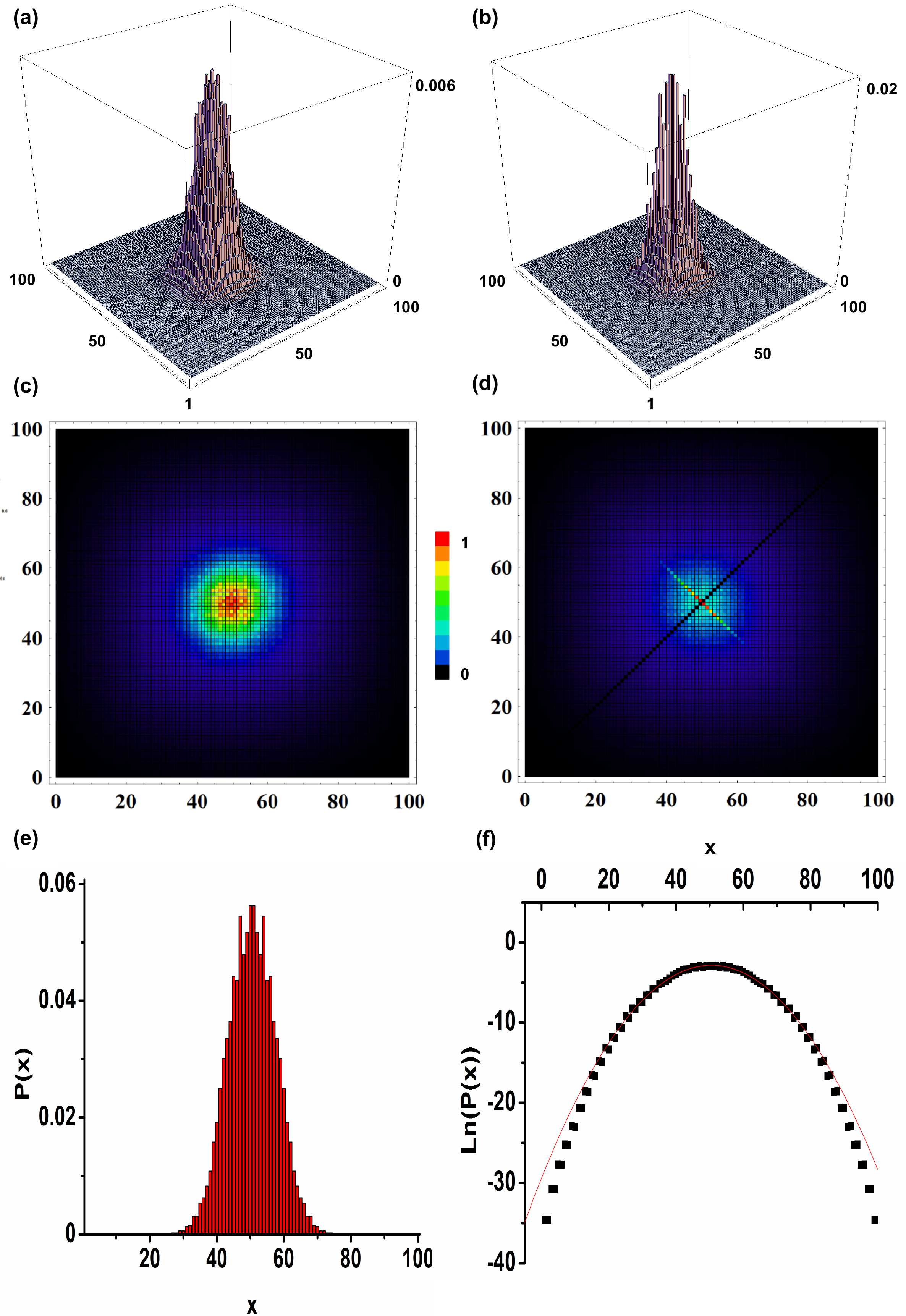}
	\caption{Modes probability distributions (a,b) and associated density plots (b,c) of bosonic (left) and fermionic (right) two-particle states in the case of dynamic disorder. At $t=0$, two particles are placed at two neighboring sites in the center of the lattice. The distribution is calculated after $50$ steps and averaged over $100$ configurations of dynamic disorder. Both bosons and fermions localize nearby their initial position with a typical binomial distribution profile characteristic of classical random walk but the dependence on the symmetry of the initial state is still observable. {(e) Marginal distribution $P^{sym}(x)$ of Eq. (\ref{marginal}) obtained by summing over the columns of plots (c,d).}  Parabolic fit in semilog scale (f).}
\label{fig:qwdinamico}
\end{figure}

{In this section we consider the case of a two-particle quantum walk in the space correlated dynamic disorder
simulating the transition from quantum walk to classical random walk for bosons and fermions.\\
{The evolution} is related to degrees of freedom defining the system as external fields, temperature, pressure or doping \citep{Lifshits1988}. 
Dynamic disorder leads to a decoherence} effect induced by the dynamic degrees of freedom: decoherence is the cost of extracting knowledge about the state of a system \citep{Kendon2006}. By turning on an interaction between the quantum system and the environment, the fluctuations in the environment lead to a reduction of the coherence in the quantum system \citep{Kendon2006}. Thus decoherence plays a crucial role in the transition from quantum to classical world. The controlled introduction of decoherence enables a detailed comprehension of its effects on the system dynamics, enabling the simulation of biological phenomena which exploit these features \citep{Lloyd2008}.
\\
\indent
The impact of decoherence on quantum walks has been investigated both numerically and experimentally, in various settings, mostly in one- \citep{Kendon2006,Silberhorn2011,cres13npo} or two-dimensional systems \citep{schr12sci} using repeated measurements \citep{Kendon2006,Kempe2003,M.A.Nielsen2000} or topological noise by quantum coin phase shift operations \citep{Silberhorn2011,Chandrashekar2011,Kendon2006,cres13npo}. In this case the time-dependent coin operator reads
\begin{equation}
C(t)=\frac{1}{\sqrt{2}}\left(
	\begin{array}{lr}
	e^{i\phi_{L}(t)} &0 \\
	0 &e^{i\phi_{R}(t)}
	\end{array}
	\right)\left(
\begin{array}{lr}
1 &1 \\
1 &-1
\end{array}
\right).
\end{equation}
By assigning different quantum coin operators at every step of the walk while retaining the same at every spatial site $(\phi_{L}(t)=\phi_{R}(t))$, thus eliminating position dependent phase correlations, the complete evolution will be expressed as $C(t)C(t-1)\ldots C(2)C(1)$ \citep{Silberhorn2010,Chandrashekar2011}.\\
Decoherence appears as a consequence of the dynamically varying phase suffered by the quantum particle during its evolution. As a result, the photon undergoes a classical random walk, revealing a binomial probability distribution \citep{Silberhorn2011,Kendon2006}. In contrast to the previous case, the spatial profile of the wave packet in Fig. \ref{fig:qwdinamico} shows a parabolic shape in the semilog scale, confirming a Gaussian distribution profile.\\
\indent
It is important to highlight that, by introducing decoherence in quantum walks by quantum coin phase-shift operations, the system dynamics is still unitary because phase-shift operator is unitary. Hence the decoherence effect is caused only by an interference effect that mixes the position distribution of the quantum walk to a uniform distribution essentially as in the classical case. Therefore, in presence of dynamic disorder quantum walk is still strongly dependent on the symmetry of the input state. Quantum particles lose their quantum waveform (but not their quantum properties) given by the delocalized spatial distribution profile undergoing a classical particle distribution profile, exhibiting the wave-particle dualism.
\section{Uncorrelated dynamic disorder and Anderson transition}
\label{sec:fluctuatingdisorderandandersontransition}
\begin{figure}[ht]
	\centering
		\includegraphics[width=\columnwidth]{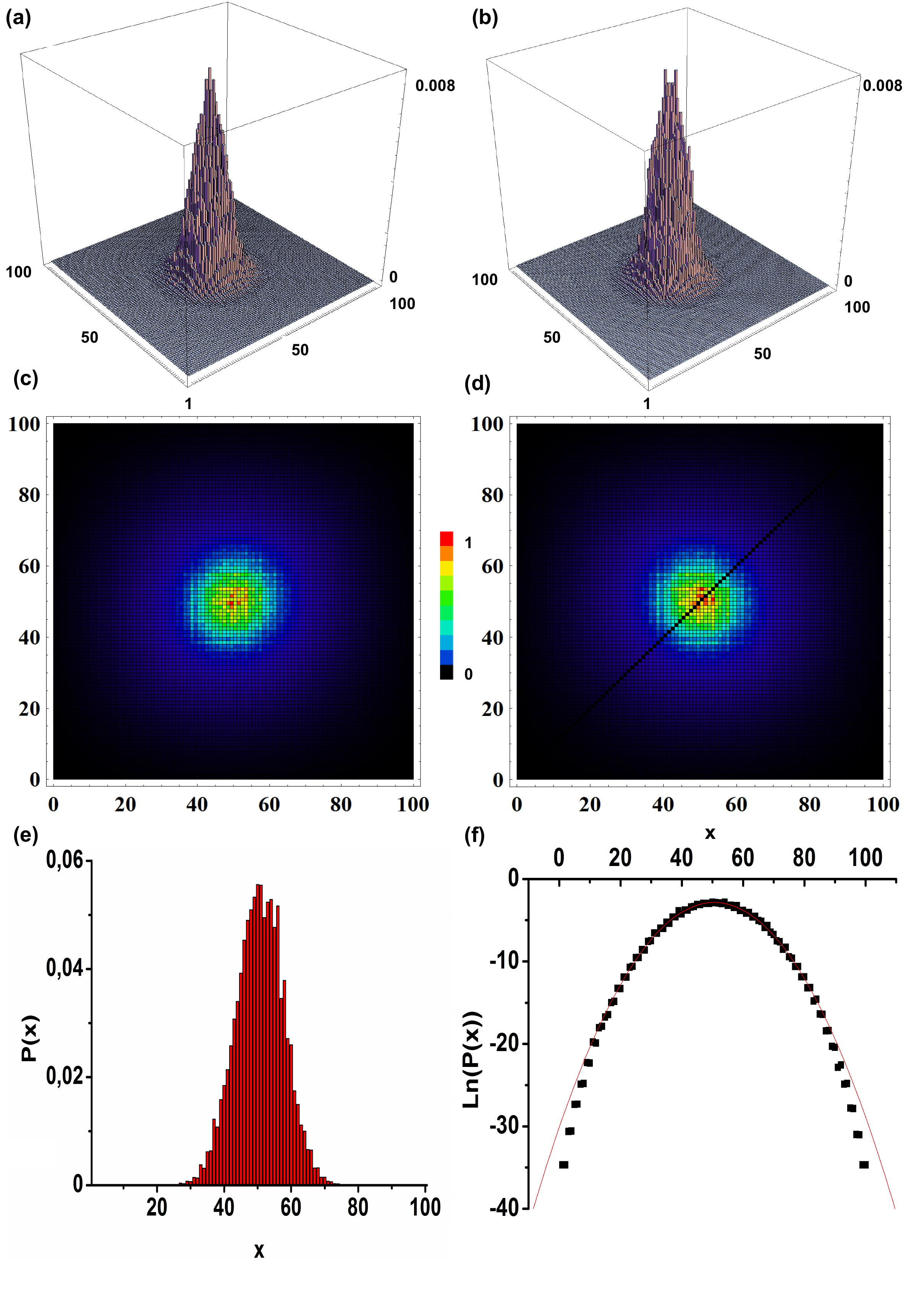}
	\caption{Modes probability distributions (a,b) and associated density plots (b,c) of bosonic (left) and fermionic (right) two-particle states in the case of fluctuating disorder. At $t=0$, two particles are placed at two neighboring sites in the center of the lattice. The distribution is calculated after $50$ steps and averaged over $100$ configurations of disorder. Space-correlated dynamic disorder competes with static disorder extending quantum states and forbidding localization. The result is that both bosons and fermions distributions become Gaussian and the dependence on the symmetry of the initial states still remains. (e) Marginal distribution $P^{sym}(x)$ of Eq. (\ref{marginal}) obtained by summing over the columns of plots (c,d). Parabolic fits in semilog scale (f).}
	\label{fig:qwfluttuazioni}
\end{figure}

A particle moving in a spatially-disordered time-independent potential can exhibit Anderson localization.  At the same time, it is also known that, if the disordered potential is also fluctuating in time, localization is lost and transport is restored. This is the so-called Anderson transition, observed in a great variety of experimental conditions, { from electro-magnetic waves propagating in strongly disordered dielectric structure such as, doped semiconductors and amorphous systems \citep{Anderson1958,Castellani1983,Lee1985,Mott1967}. This transition can be obtained by changing temperature, pressure, doping \citep{Mott1967} or magnetization \citep{Lee1985}.}
 Over the years, several different mechanisms have been proposed for the breakdown of Anderson localization due to temporal fluctuations of the potential \citep{Aguer2010,Havlin2002,Krivolapov2011}. Mott considered the effect of phonons at low temperatures, and argued that this gives rise to a diffusive motion known as \textit{variable-range hopping conductivity}. Mott also considered the effects of a weak AC field, and suggested that a resonant interaction dominates the low-frequency response \citep{Mott1967,Krivolapov2011}.
\\
\indent
By combining the two previous disordered scenarios, we may simulate a symmetric or an antisymmetric system in which localized states become extended states with time due to decoherence effects. By randomly assigning different quantum coin operators for each lattice site and changing these operators during each step of the evolution, we induce a spatio-temporal disorder or \textit{fluctuating disorder} \citep{Chandrashekar2011}. Therefore the quantum coin operator of the $t^{th}$ step of the dynamics can be expressed as
\begin{equation}
C(x,t)=\frac{1}{\sqrt{2}}\left(
	\begin{array}{lr}
	e^{i\phi_{L}(x,t)} &0 \\
	0 &e^{i\phi_{R}(x,t)}
	\end{array}
	\right)\left(
\begin{array}{lr}
1 &1 \\
1 &-1
\end{array}
\right)
\end{equation}
where the value of $\phi_{L,R}(x,t)$ is randomly chosen at every step $t$ and at each site $x$. Thus the complete evolution of the walk will be given by $C(x,t)C(x,t-1)\ldots C(x,2)C(x,1)$ \citep{Chandrashekar2011}.\\
\indent
{This scenario leads to distributions similar to those obtained in presence of space-correlated dynamic disorder (compare Figs. \ref{fig:qwdinamico} and \ref{fig:qwfluttuazioni}). Fig. \ref{fig:qwfluttuazioni} shows that with an increasing number of steps the dynamic variation of the phase competes with the localization effect given by the static variation of the phase, displaying a Gaussian probability distribution. In the space-correlated dynamic disordered scenario, we have shown that particles started in a disorder-free configuration with extended distributions of quantum states and the effect of disorder was to shrink the wave functions displaying a Gaussian profile. Here, input states first localize in an Anderson peak owing to static disorder, and only after the distributions become binomial due to decoherence: static disorder has a priority upon dynamic disorder because the former acts on different spatial sites of the same temporal step, while the latter's action is the same for all the site of a given step. Therefore, dynamic disorder is a global effect on the walk and it is slower than the static effect, as we will show in the next section.\\
Also in this case, the quantum walk still has a strong dependence  from the symmetry of the input state, as we can notice in Fig. \ref{fig:qwfluttuazioni}. Unlike Fig. \ref{fig:qwdinamico}d, the fermionic distribution in Fig. \ref{fig:qwfluttuazioni}d has a Gaussian spatial profile more pronounced and very similar to the bosonic wave packet. This is due to the fact that, as we said, input states localize (Fig. \ref{fig:qwstatico}) before becoming extended.
}

\section{Disorder strength}
\label{sec:Disorderstrength}
\begin{figure}[ht]
	\centering
		\includegraphics[width=\columnwidth]{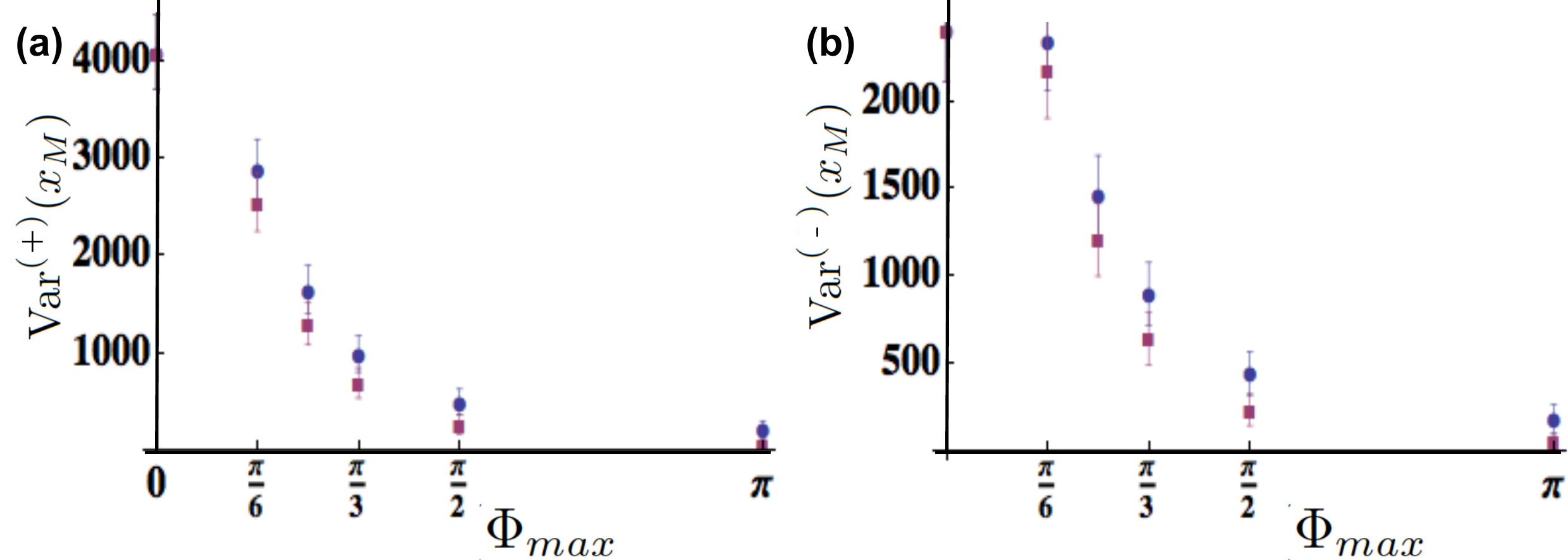}
	\caption{Transition of the mean position variance $\text{Var}^{(\pm)}(x_M)$  of Eq. (\ref{eq:variance}) of symmetric (a) and antisymmetric (b) states from ballistic quantum walk to diffusive or subdiffusive evolution after $100$ steps due to dynamic (blue dots) and static (purple squares) disorder with increasing disorder strength $\Phi_{max}$. Every data point is averaged over $100$ configurations of disorder and the error bars represent the standard deviation. The black solid line marks the variance of the classical random walk (RW).}
	\label{fig:forza}
\end{figure}

\begin{figure}[ht]
	\centering
		\includegraphics[width=\columnwidth]{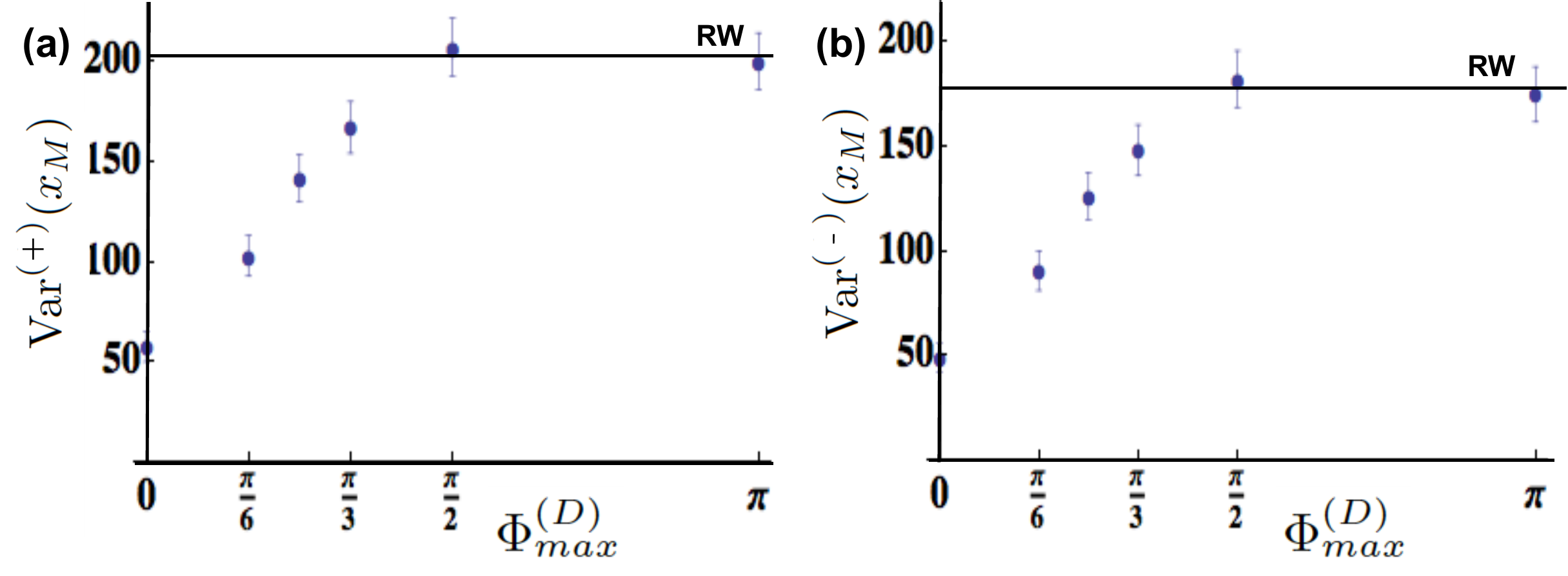}
	\caption{Transition of the  mean position variance $\text{Var}^{(\pm)}(x_M)$  of Eq. (\ref{eq:variance}) of symmetric (a) and antisymmetric (b) states from subdiffusive quantum walk to diffusive evolution after $100$ steps due to dynamic disorder (blue dots) with increasing dynamic disorder strength $\Phi_{max}^{(D)}$ once fixed the static disorder strength $\Phi_{max}^{(S)}=\pi$. The mobility edge is $\Phi_{max}^{(D)}\cong\pi/2$. Every data point is averaged over $100$ configurations of disorder and the error bars represent the standard deviation. The black solid line marks the variance of the classical random walk (RW).}
	\label{fig:mobility}
\end{figure}

When a disordered scenario is considered, the degree of disorder needs to be characterized quantitatively. This physical quantity can be addressed by taking into account the disorder strength which is determined by {$\Phi_{max}$, i.e. the maximal applied phase shift which defines the uniform interval $\left[0,\,\Phi_{max}\right]$ from which $\left\{\phi_{L}(x),\,\phi_{R}(x)\right\}$ are randomly chosen} \citep{Silberhorn2011}.\\ 
\indent
The stepwise increase of the disorder strength $\Phi_{max}$ enables the controlled transition of the system from the ballistic evolution of the disorder-free quantum walk, towards the diffusive (subdiffusive) evolution in a scenario with dynamic (static) disorder, as shown in Fig. \ref{fig:forza}.\\
{To this purpose we characterized the resulting expansion profile by the two-particle distribution variance $\text{Var}^{(\pm)}(x_M)$ of twice their mean position $x_M = x+y$, i.e.
\begin{eqnarray}
\label{eq:variance}
\text{Var}^{(\pm)}(x_M)=&\sum_{x,y=1}^N(x+y)^2P^{(\pm,sym)}(x,y)-\\
&[\sum_{x,y=1}^N(x+y)P^{(\pm,sym)}(x,y)]^2.\nonumber
\end{eqnarray}
Clearly, this quantity
will result in different behaviors for the symmetric state (related to bosonic evolution) and the antisymmetric one (related to fermionic evolution), the exact connection with the particle statistics being detailed in Eq. (A12). Moreover it will be strongly affected by the presence of disorder.\\
In Fig. \ref{fig:forza} we report the variance (\ref{eq:variance}) calculated for 100-step quantum walks in presence of static (red squares) and space-dependent dynamic (blue dots) disorder, with different values of the maximum disorder strength, averaged on 100 random distributions. 
}
Without disorder $(\Phi_{max}=0)$ the ballistically spreading wavepacket shows a large expansion induced by quantum interference after $100$ steps. In a system with dynamic disorder (blue squares), decoherence reduces the expansion of the wave packet to the level of a diffusive classical particle (black solid line).\\
Differently, static disorder (red dots) leads to a stagnation of the spread due to the Anderson localization effect and hence an even smaller variance. The variance exponentially decreases as the disorder strength increases and in general as the number of steps grows. {By comparing the two behaviors reported in Fig. \ref{fig:forza} , we may notice that the presence of static disorder tends to localize the distribution faster than the space-correlated dynamic disorder, since, for each value of disorder strength, the variance in presence of space-dependent disorder is lower than the variance obtained in the dynamic-disordered scenario.} We infer from Fig. \ref{fig:forza}a that the maximal localizing interval is $\left[0,\,\pi\right]$,  irrespective of the particle statistics. This is the same interval we used in all the previous simulations. These results clearly demonstrate how the amount and kind of disorder influence the expansion of the particle wave packet.
\\
\indent
Let us now consider the variance of a quantum walk in the presence of uncorrelated dynamic disorder (blue dots in Fig. \ref{fig:mobility}) and a stepwise increase of the dynamic disorder strength $\Phi_{max}^{(D)}$ once fixed the static disorder strength to its highest value $\Phi_{max}^{(S)}=\pi$.
From Fig. \ref{fig:mobility} it is possible to rate the dynamic disorder strength necessary to extend the localized states and realize in this way a controlled Anderson transition. The mobility edge is defined as the critical value of disorder strength for which the variance reaches the level of classical random walk (black solid line). For values of $\Phi_{max}^{D}$ below the mobility edge the variance decreases because static disorder dominates, thus localizing the particle wave packet.
\section{Transport properties and fractality}
\label{sec:Transportproperties}
In this section we study how transport properties of particles are connected to the fractality of the system. 
Fractals are mathematical objects with a Hausdorff-Besicovitch dimension which is not an integer \citep{Havlin2002,Mandelbrot1977}. Fractals are best constructed in a recursive way and their limiting curve is of infinite length, although it is confined to a finite region of the plane. The best way to characterize them is by using their fractal dimension $d$. When calculating this quantity we notice the striking property of self-similarity: every fractal curve is similar to the fractal curve of a part of itself or, in other words, it is said to be a scale invariant.
\\
\indent
Classical random walks are self-similar only in a statistical sense \citep{Havlin2002}. For these objects a fractal dimension $d$ is still easily defined by the scaling of their fractal mass $M$ with their linear size $L$, $M\approx L^{d}$. As a consequence, classical random walks are statistical fractals with Hausdorff dimension $d=2$ \citep{Havlin2002}. Conversely, quantum walks are not fractals not even in statistical sense: because of their quantum nature, their fractal dimension is $d=1$.
\\
\indent
Particle transport in a lattice has been widely investigated in the framework of quantum walk \citep{Silberhorn2011,Lloyd2008,Havlin2002}. One of the most important physical parameters describing a random walk is the mean-square displacement covered by the walker once it has passed through $t$ steps. This quantity is proportional to the variance of the center of mass of Eq. (\ref{eq:variance}). In uniform Euclidean systems, the mean-square displacement of a random walker is proportional to the time, for any number of spatial dimensions agreeing with Fick's law \citep{Havlin2002}. Random walks describe the probabilistic evolution of a classical particle in a structured space resulting in a diffusive transport. In contrast, endowing the walker with quantum mechanical properties typically leads to a ballistic spread of the particle's wave function \citep{Silberhorn2011}.\\
However, in disordered systems, Fick's law is not valid in general. Rather, we may recall the localization behavior as anomalous diffusion \citep{Havlin2002}: $\left\langle\textbf{x}^{2}(t)\right\rangle\propto t^{\frac{2}{d}}$.
\begin{figure}[t!!!]
	\centering
		\includegraphics[width=\columnwidth]{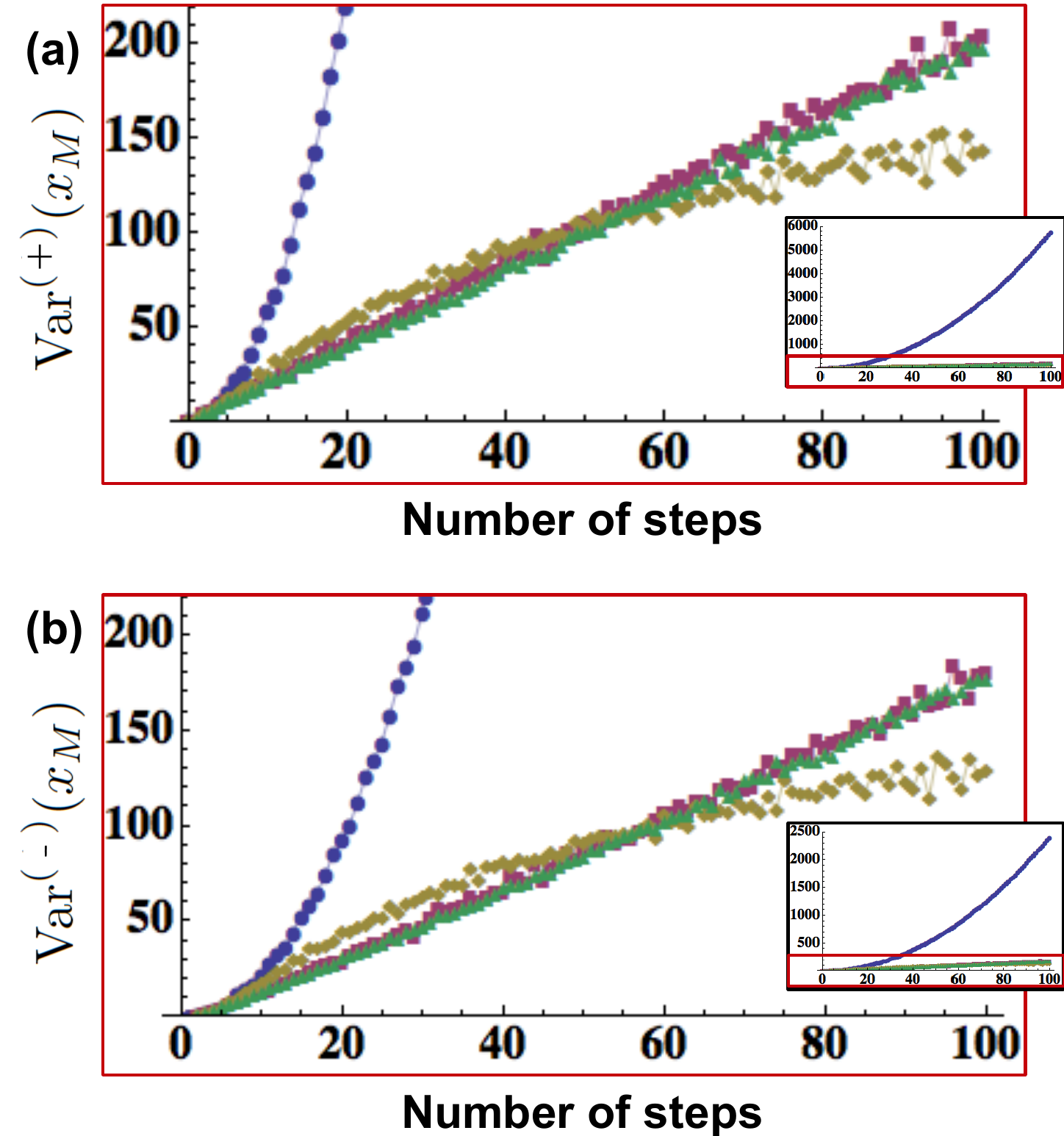}
	\caption{Trend of mean position variance $\text{Var}^{(\pm)}(x_M)$  of Eq. (\ref{eq:variance}) up to $100$ steps of bosonic (a) and fermionic (b) quantum walk in different scenarios. For disorder-free case (blue dots) we observe a ballistic spread (full behavior shown in the insets). The evolutions with space-correlated dynamic (purple squares) and uncorrelated (green triangles) disorder are clearly diffusive. Lastly, under the condition of static disorder (yellow diamond) the variance starts to saturate after few steps and the dynamics shows the onset of Anderson localization. The parameters used in simulation are equivalent to the simulational settings used for Figs.  \ref{fig:qwordinato},\ref{fig:qwstatico},\ref{fig:qwdinamico},\ref{fig:qwfluttuazioni}. Every data point for the four kinds of disordered quantum walks is averaged over $100$ configurations of disorder. Error bars have been omitted for clarity.}
	\label{fig:diffusione}
\end{figure}
Note that classical Fick diffusion with $d=2$ can be considered as an anomalous diffusion with respect to the quantum case with $d=1$.\\
The observed slowing down of the transport is caused by the delay of the diffusing particles in the dangling ends, bottlenecks and backbends existing in the disordered structure. Examples of disordered systems for which anomalous diffusion has been observed are percolation clusters \citep{Gennes1976,Mitescu1977}, fractal lattices \cite{Havlin2002}, classical and quantum random walks \cite{Silberhorn2011,cres13npo}.
\\
\indent
In Fig. \ref{fig:diffusione} we report step by step the variance $\text{Var}^{(\pm)}(x_M)$  of Eq. (\ref{eq:variance}) of the center of mass in all the previous four quantum walk scenarios (the disorder strength is maximum), for symmetric and antisymmetric input states.
\\
First we notice that, due to the symmetry of the states, bosons diffuse faster than fermions reaching higher variance values. Moreover, by computing polynomial fits of these curves we obtain the correct trends of the mean-squared displacement and the fractal dimensions, for the  two-particle quantum walk with bosonic and fermionic input states. In the disorder-free case the quantum walk has fractal dimension $d\approx1$; therefore the diffusion, characterized here by the positional probability distribution variance, is ballistic $\text{Var}^{(\pm)}(x_M,t)\propto t^{2}$. In presence of dynamic and fluctuating disorder, the fractal dimension is $d\approx2$ therefore the motion is diffusive $\text{Var}^{(\pm)}(x_M,t)\propto t$. Finally, in the case of static disordered quantum walks the fractal dimension is $d\approx3.4$ and the trend is subdiffusive $\text{Var}^{(\pm)}(x_M,t)\propto t^{0.6}$, thus leading to particle stagnation.   
{
\section{Shannon entropy and mutual information}

\label{sec:Shannon entropy}

An alternative method to measure the position fluctuation with the variance is provided by the joint Shannon entropy of the two walkers position probability distribution $P^{(\pm, sym)}(x,y)$ which via Eq.(\ref{eq:variance}) describe the statistics in the Bosonic (resp. Fermionic) case. This is
\begin{equation}
H(X,Y)\equiv -\sum_{x,y}P^{(\pm, sym)}(x,y)\log_{2}P^{(\pm, sym)}(x,y).
\label{eq:entropy}
\end{equation}
\begin{figure}[b]
	\centering
		\includegraphics[width=\columnwidth]{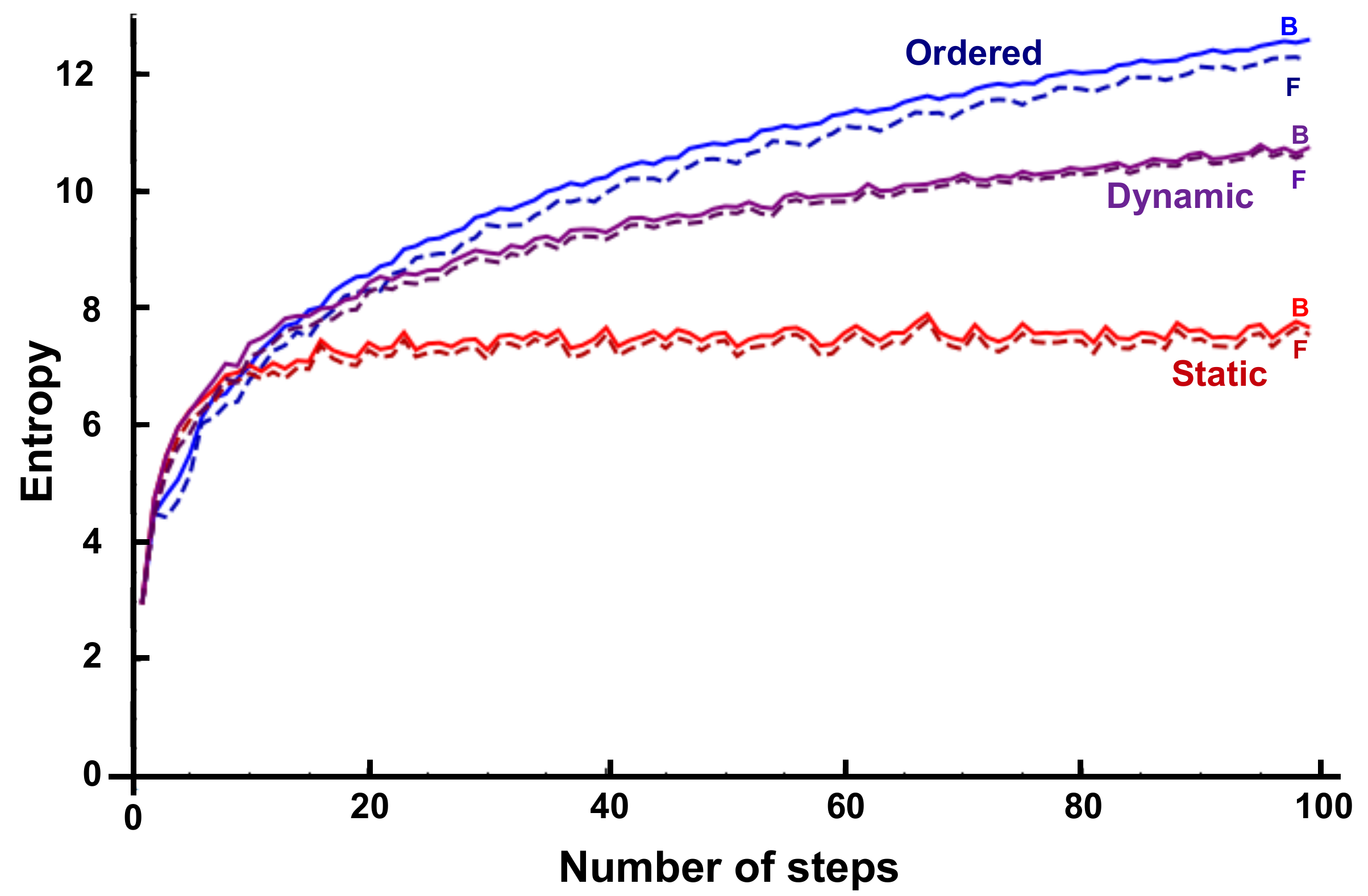}
	\caption{Joint Shannon entropy (\ref{eq:entropy}) up to $100$ steps of symmetric (solid curves) and antisymmetric (dashed curves) quantum walk in different conditions. We may observe how the increase of entropy is slowed when passing from the ordered (blu curves) to the disordered case, saturating in the case of static disorder (red curve). For each type of disorder the distribution $P^{(-,sym)}(x,y)$ entropy is always lower than the entropy of $P^{(+,sym)}(x,y)$ one. Each data point for the three kinds of disordered quantum walks is averaged over $50$ configurations of disorder. Error bars have been omitted for clarity.}
	\label{fig:entropia}
\end{figure}
{}This quantity varies with the number of steps in a similar way for symmetric and antisymmetric particles, however values of $H(X,Y)$ obtained at each step depend on the particle statistics. In particular we may observe from Fig. \ref{fig:entropia} that the distribution $P^{(-,sym)}(x,y)$ share a joint entropy lower than the one shared by the distribution $P^{(+,sym)}(x,y)$, and this happens in presence of any kind of disorder. Shannon entropy, that we may consider as the amount of information gained by increasing the number of steps of the QW, depends on the type of disorder the walker undergoes during the walk.

\begin{figure}[b]
	\centering
		\includegraphics[width=\columnwidth]{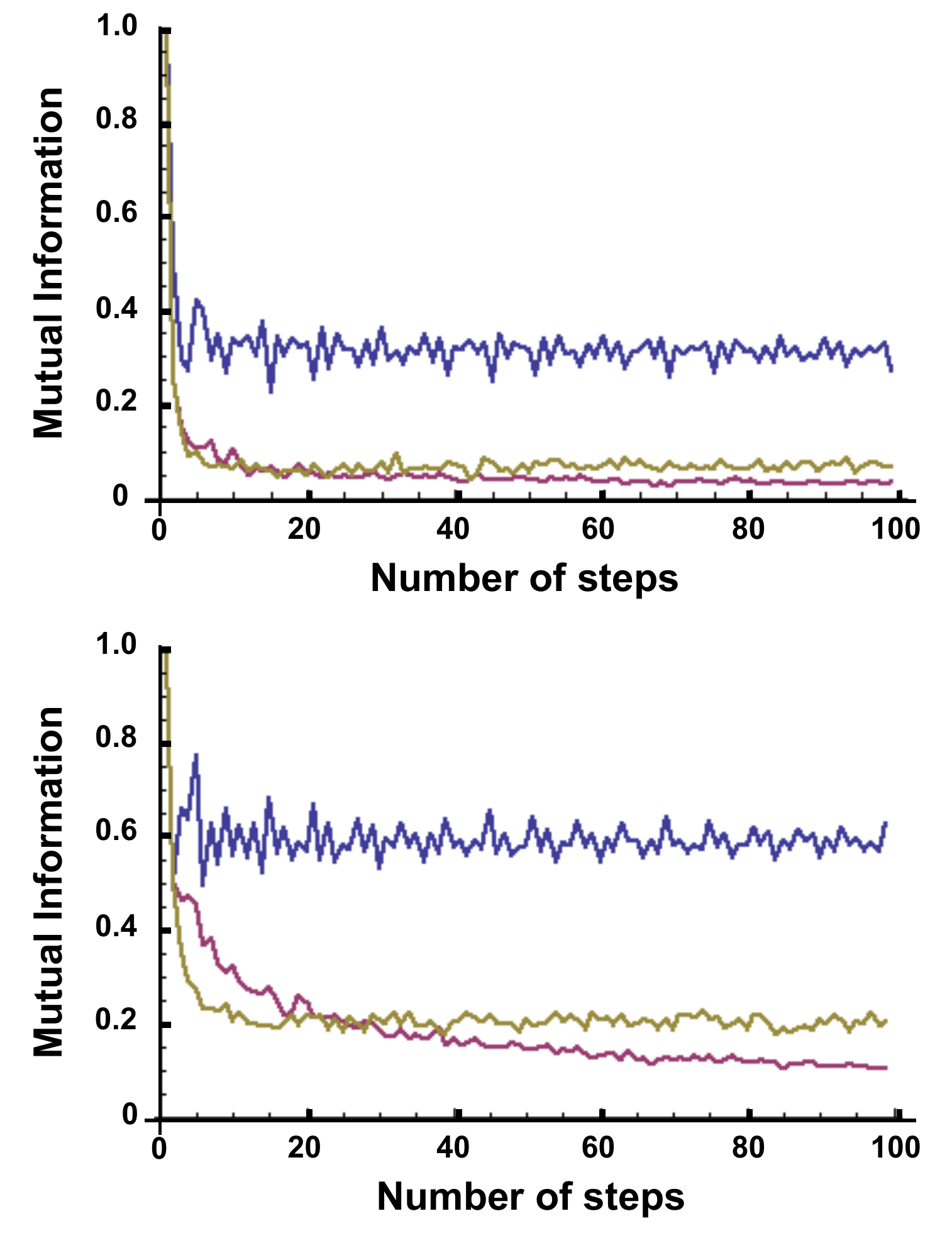}
	\caption{Mutual information (\ref{eq:MI}) up to $100$ steps of the quantum walk for (a) the distribution $P^{(+,sym)}(x,y)$ and (b) the distribution $P^{(-,sym)}(x,y)$ in the disorder-free case (blue line), dynamic disordered (purple line) and static disordered case (yellow line). Mutual information decreases as the number of steps increase. At each step the mutual information for bosons is lower than the one for fermions. Every data point for the three kinds of disordered quantum walks is averaged over $50$ configurations of disorder. Error bars have been omitted for clarity.}
	\label{fig:mutualinformation}
\end{figure}
As reported in Fig. \ref{fig:mutualinformation} we also calculated another quantity, the mutual information. The mutual information content of $X$  and $Y$ measures the amount of common information about the spatial position of the two particles and its expression reads:
\begin{eqnarray}
\label{eq:MI}
	I^{(\pm)}(X:Y)&=& H(X)+H(Y) - H^{(\pm)}(X,Y)\\
	&=& 2 H(X) - H^{(\pm)}(X,Y),\nonumber
\end{eqnarray}
where $H$ is the Shannon entropy associated with the single particle marginal (A10) [notice that it does not depend upon the symmetry of the input state, see Appendix A].

We may observe in Fig. \ref{fig:mutualinformation} that this quantity reaches an asymptotic value in the disorder-free case and in presence of static disorder, while it slowly decreases in presence of dynamic disorder. In presence of this type of disorder, indeed, the distribution tends to the classical one, so we may imagine that one particle loses information about the position of the other one since they tend to behave like independent walkers.
Both the symmetric and the antisymmetric distributions exhibit this behavior, however symmetric particle systems share a mutual information lower than the one shared by antisymmetric particles. This may be naively understood by observing that, since the antisymmetric distribution is related to fermionic behavior, due to Pauli exclusion principle, fermions can't occupy all possible two-particle positions: this corresponds to a residual mutual knowledge between the two particles.}

\section{Conclusions}
\label{sec:conclusions}
In this paper, we presented a theoretical analysis of two non-interacting bosons and fermions travelling in a discrete-time quantum walk.
By varying the parameters of the system, transport properties have been studied. Then the interplay between quantum coherence and the presence of dephasing disorder has been investigated, with particular attention to entanglement and disorder-assisted transport effects. By introducing suitable static disorder in the walk we have simulated the absence of diffusion in a periodic lattice. In these conditions, the onset of Anderson localization has been observed. Furthermore, decoherence deriving from dynamic disorder has been investigated, simulating the transition between quantum and classical world.\\
Finally we have presented how fluctuating disorder can lead to a deeper comprehension of the Anderson transition.
\section{Acknowledgement}
\label{sec:ringraziamenti}
This project was supported by FIRB- Futuro in Ricerca HYTEQ, PRIN 2009 and ERC: 3D-QUEST.

\newpage
{
\appendix
{{
\section{Polarization and Statistics}
Given  the state (\ref{INPUT2}) the probability of detecting a photon in position $x$ and the other in position
$y$ irrespectively of their polarization writes
\begin{widetext}
\begin{eqnarray}
 P^{(\pm)}(x,y,t )  = \left\{  \begin{array}{ll} 
 \big| \langle x , H | \langle y, V | \Psi^{(\pm)} (t) \rangle \big|^2 + \big| \langle y , V | \langle x, H |\Psi^{(\pm)}(t) \rangle \big|^2 & i > j\;, \\\\
\big| \langle x , H | \langle x, V | \Psi^{(\pm)} (t)\rangle \big|^2  & x= y\;,
\end{array} 
\right. 
 \end{eqnarray} 
 where $|x, H\rangle$ and $|x,V\rangle$ describe  a photon emerging from the output port $x$ of the
 setup  with horizontal and vertical polarization respectively. 
 Explicitly this is
 \begin{eqnarray}
 P^{(\pm)}(x,y,t)= \left\{  \begin{array}{ll} 
\big| \psi_A(x,t) \psi_B(y,t)  \pm \psi_A(x,t) \psi_B(y,t) \big|^2 &  \mbox{for $x > y$} \;, \\\\
\left\{ 
\begin{array}{ll} 
2\; \big| \psi_A(x,t) \psi_B(y,t) \big|^2 &  \mbox{for $(+)$}\\\\
0 & \mbox{for $(-)$}
\end{array} 
\right.
 & \mbox{for $x = y$}\;, 
\end{array} 
\right. 
 \end{eqnarray} 
  \end{widetext} 
 where for $C=A,B$ we introduced the single particle amplitude probabilities, 
 \begin{eqnarray}
\psi_{C}(x,t) = \langle x | \psi_C(t)\rangle\;.
\end{eqnarray} 
(we stress that  to avoid double counting,  $P^{(\pm)}(x,y,t)$ is defined {\it only} for $x\geq y$).

The probabilities defined above correspond to the probabilities one would get if the particles were indistinguishable and obeyed to the Bosonic/Fermionic statistics. This is a consequence of two facts: i) while in  Eq.~(\ref{INPUT2})   the two particles are distinguishable in terms of their  polarization degree of freedom, the measurement  we consider
is transparent with   respect to this degree of freedom (we do not distinguish whether the emerging photon is $H$ or $V$); ii) the vector $|\Psi^{(\pm)} \rangle$ is symmetric (resp. anti-symmetric) for particle (i.e. polarization) exchange.
To see this explicit  observe that in first quantization, having identified the polarization with the particle indexes, we have that $|\Psi^{(+)}(t)\rangle$  define a proper state  of two  Bosons.
  The probability of detecting one of the two particles in position  $x$ and the other   in $y(<x)$, can then be obtained by projecting  $|\Psi^{(+)}(t)\rangle$ in the Bosonic state which represents such final configuration  (i.e. the symmetric vector $(| x,H \rangle |y,V\rangle  + | y,H\rangle | x, V\rangle)/{\sqrt{2}}$),  
\begin{eqnarray}
&&P^{(BOS)}(x,y) =\Big|  \left[ \frac{  \langle x, H |\langle y, V|+  \langle y, H|\langle x, V| }{\sqrt{2}}\right]
\; | \Psi^{(+)}\rangle
 \Big|^2  \nonumber \\
 &=& \big| \psi_A(x) \psi_B(y)  +  \psi_A(x) \psi_B(x) \big|^2 = P^{(+)}(x,y)\;, \end{eqnarray} 
(hereafter the time has been neglected for easy of notation).
Similarly the probability of detecting both particles in $x$ is given by
\begin{eqnarray}
P^{(BOS)}(x,x) &=&\Big| \langle x, H|\langle x,V | \; | \Psi^{(+)} \rangle 
 \Big|^2  \\
 \nonumber 
  &=& 2\; \big| \psi_A(x) \psi_B(y) \big|^2= P^{(+)}(x,x)\;. \nonumber 
\end{eqnarray} 
The same conclusions applies for Fermions. Indeed in this case one has
\begin{eqnarray}
&& P^{(FER)} (x,y)=\Big|  \frac{ \left[ \langle x, H |\langle y,V |-  \langle y,H |\langle x,V| \right]}{\sqrt{2}}
\; | \Psi^{(-)} \rangle
 \Big|^2  \nonumber \\
 &=& \big| \psi_A(x) \psi_B(y)  -  \psi_A(y) \psi_B(x) \big|^2 = P^{(-)}(x,y)\;, \end{eqnarray} 
for $x>y$, while of course 
\begin{eqnarray}
P^{(FER)}(x,x) =0 = P^{(-)}(x,x)\;.
\end{eqnarray}

A compact way to express the above quantities is by means of the 
 symmetric version of the distributions $P^{(\pm)}(x,y)$, i.e. the quantities
\begin{eqnarray} \label{defiSYM}
P^{(\pm, sym)}(x,y) &=& P^{(\pm, sym)}(y,x) \\ \nonumber 
&=& \frac{ \big| \psi_A(x) \psi_B(y)  \pm \psi_A(y) \phi_B(x) \big|^2}{2}\;. 
\end{eqnarray} 
The functions $P^{(\pm, sym)}(x,y)$  are normalized when integrated over the whole domain of $x$ and $y$, i.e. 
\begin{eqnarray}
&& \sum_{x,y} P^{(\pm, sym)}(x,y) = \sum_{x,y} \frac{ \big| \psi_A(x) \psi_B(y)  \pm \psi_A(y) \psi_A(y) \big|^2}{2} \nonumber \\
& &\quad = \langle \psi_A | \psi_A \rangle \langle \psi_B | \psi_B \rangle \pm  |\langle \psi_A | \psi_B \rangle|^2 =1 \;.
\label{NORMA}
\end{eqnarray} 
and  posses identical marginals, i.e.
\begin{eqnarray}
&& P^{(sym)}(x) = \sum_{y} P^{(\pm, sym)}(x,y)
\nonumber \\
&& =\frac{ \big| \psi_A(x) \big|^2  + \big| \psi_B(x) \big|^2}{2}  \pm \mbox{Re} [ \psi_A(x) \psi_B(x)^* 
\langle \psi_B | \psi_A \rangle] \nonumber \\
&&= \frac{ \big| \psi_A(x) \big|^2  + \big| \psi_B(x) \big|^2}{2}   \label{marginal}\;,
\end{eqnarray} 
(notice  that in writing the last identities of Eq.~(\ref{NORMA}) and (\ref{marginal})  we explicitly used the fact that $|\psi_A\rangle$ and $|\psi_B\rangle$ are orthonormal states: these vectors are in fact associated with two single particle trajectories entering the system from two distinct ports of the interferometer).
Accordingly we can write 
\begin{eqnarray}
P^{(BOS)}(x,y) &=& 2 P^{(+, sym)}(x,y) \qquad   \mbox{for $x>y$}\;, \nonumber \\
P^{(BOS)}(x,x) &=& P^{(+, sym)}(x,x) \qquad   \mbox{for $x=y$}\;, \nonumber \\
P^{(FER)}(x,y) &=& 2 P^{(-, sym)}(x,y) \qquad   \mbox{for $x>y$}\;, \nonumber \\
P^{(BOS)}(x,x) &=& P^{(-, sym)}(x,x)=0 \quad   \mbox{for $x=y$}\;.  \nonumber 
\\ 
\end{eqnarray}
In particular 
 the expectation value of any two particle observable $\Omega$ which is symmetric under particle exchange can be expressed in terms of $P^{(\pm, sym)}(x,y)$. Indeed, indicating with $\Omega(x,y)= \Omega(y,x)$ the  spatial representation of  $\Omega$, the following identity holds:
\begin{eqnarray}
\langle \Omega^{(BOS)}\rangle &=& \sum_{x\geq y} P^{(BOS)}(x,y) \Omega(x,y) \nonumber  \\
&=& \sum_{x,y} P^{(+,sym)}(x,y) \Omega(x,y) \;,\nonumber \\
\langle \Omega^{(FER)}\rangle &=& \sum_{x\geq y} P^{(FER)}(x,y) \Omega(x,y) \nonumber  \\
&=& \sum_{x,y} P^{(-,sym)}(x,y) \Omega(x,y)\;.
\end{eqnarray}

\subsection{Single particle detection probabilities}
From the above equation one can easily compute the
probability of finding a particle in position $x$. In particular for Bosons one may introduce the probability $P_{>1}^{(BOS)}(x)$ of finding {\it at least} one
particle in position $x$ and the  probability $P_{1}^{(BOS)}(x)$
of having {\it exactly}  one particle in position $x$. These quantities in general differ and can be expressed as 
\begin{eqnarray}
P_{>1}^{(BOS)}(x)&=& \sum_{y (<x)} P^{(BOS)} (x,y) + 
\sum_{y (>x)} P^{(BOS)} (y,x) \nonumber \\
&& +P^{(BOS)} (x,x)  \nonumber \\ \nonumber 
&=& 2 \sum_{y} P^{(+, sym)}(x,y) - P^{(+, sym)}(x,x)
 \\
&=& 2  P^{(sym)}(x) - P^{(+, sym)}(x,x)\;, \\
P_{1}^{(BOS)}(x) &=& \sum_{y (<x)} P^{(BOS)} (x,y) + 
\sum_{y (>x)} P^{(BOS)} (y,x) \nonumber \\
& =& 2  P^{(sym)}(x) -  2 P^{(+, sym)}(x,x)\;.
\end{eqnarray} 
We stress that neither $P_{>1}^{(BOS)}(x)$  nor $P_{1}^{(BOS)}(x)$ coincide with the marginal distribution $P^{(sym)}(x)$ of (\ref{marginal}). In particular, differently from the latter, neither  
$P_{>1}^{(BOS)}(x)$  nor $P_{1}^{(BOS)}(x)$ are necessarily normalized to $1$ when summing over $x$ 
(this is due to the fact that when summing over $x$ we
are unavoidably including double counting of events). 
In the Fermionic case $P_{>1}^{(FER)}(x)$ and 
$P_{1}^{(FER)}(x)$ coincides due to Pauli exclusion principle. In this case we have 
\begin{eqnarray}
P_{1}^{(FER)}(x)&=& \sum_{y (<x)} P^{(FER)} (x,y) + 
\sum_{y (>x)} P^{(FER)} (y,x) \nonumber \\
&=& 2 \sum_{y} P^{(-, sym)}(x,y) = 2  P^{(sym)}(x)\;,
 \end{eqnarray}
which up to a constant normalization factor coincides
with 
the marginal of the symmetric distribution $P^{(-, sym)}(x,y)$. }}

\bibliographystyle{plainnat}
\bibliography{Arxive_De_Nicola}

\end{document}